\documentclass[12pt]{iopart} 
\usepackage{graphicx}
\usepackage{bm}

\begin{document}
  
\title[Mechanical properties of amorphous nanosprings]
{Mechanical properties of amorphous nanosprings}

\author{Alexandre F. da Fonseca$^1$, C. P. Malta$^1$ and D. S. 
Galv\~{a}o$^2$}

\address{$^1$ Instituto de F\'{\i}sica, Universidade de S\~ao
Paulo, Caixa Postal 66318, 05315-970, S\~ao Paulo, Brazil}

\address{$^2$ Instituto de F\'{\i}sica `Gleb Wataghin',
Universidade Estadual de Campinas, Unicamp 13083-970, Campinas, SP,
Brazil}   

\eads{\mailto{afonseca@if.usp.br}, \mailto{coraci@if.usp.br}, 
\mailto{galvao@ifi.unicamp.br}}

\begin{abstract} 

Helical amorphous nanosprings have attracted particular interest due
to their special mechanical properties. In this work we present a
simple model, within the framework of the Kirchhoff rod model, to
investigate the structural properties of nanosprings having asymmetric
cross section. We have derived expressions that can be used to obtain
the Young's modulus and Poisson's ratio of the nanospring material
composite. We also address the importance of the presence of a
catalyst in the growth process of amorphous nanosprings in terms of
the stability of helical rods.

\end{abstract} 
 
  

\submitto{\NT}

\maketitle

\section{Introduction}  

Helical nanowires, or simply {\it nanosprings}, are particularly
interesting one-dimensional nanostructures~\cite{mc1} due to their
special periodic and elastic properties. Examples of such structures
are quasi-nanosprings~\cite{tang}, helical crystalline
nanowires~\cite{zhangw,ame,wang} and amorphous
nanosprings~\cite{mc2,zhang,mc3}.

Volodin {\it et al}~\cite{volodin} have studied the elastic properties
of helix-shaped nanotubes using Atomic Force Microscopy (AFM). They
used a circular beam approximation to model the elastic response of a
single winding of coiled nanotube. Recently, Chen {\it et
al}~\cite{chen} measured the spring constant of carbon nanocoils and
used a classical approach which relates the spring constant to the
shear modulus of the composite material. 

In this paper, we present a model that can be used by experimentalists
to determine the elastic properties of different amorphous
nanosprings. Our calculations are based on the Kirchhoff rod
model~\cite{kirchhoff} that provides a framework to study statics and
dynamics of thin elastic filaments. The static Kirchhoff equations
will be used to derive expressions for the Young's modulus and the
Poisson's ratio of the nanospring material composite. The nanowire
here is assumed to have elliptic cross section, so the present work
constitutes an extension of our previous works~\cite{douglas,fonseca5}
for nanowires with circular cross-section.

It is known that the synthesis of amorphous nanowires and nanosprings
requires the presence of a metallic catalyst. Following the
vapor-liquid-solid (VLS) growth model, known since 1964~\cite{ellis}
for whisker formation, a liquid droplet of a metal absorbs a given
material from the surrounding vapor, and after super-saturation of the
absorbed material within the droplet, the excess material precipitates
at the liquid-solid interface forming the nanowire beneath the
metallic droplet.

McIlroy {\it et al}~\cite{mc1,mc2} developed a modified VLS growth
model to explain the formation of amorphous nanosprings based on the
interactions between the metallic catalyst and the nanowire. The
interesting feature is that the modified VLS growth model~\cite{mc1}
does not depend on the composite material of the nanospring and,
therefore, it can be applied to any type of amorphous nanosprings.

Here, we will take advantage of the modified VLS growth model proposed
by McIlroy {\it et al}~\cite{mc1} to analyse an important mechanical
consequence of the presence of the metallic catalyst in the growth
process of amorphous nanosprings. We show that the asymmetric growth
driven by the metallic catalyst provides the nanowire with a {\it
helical intrinsic curvature} which is required to maintain it
dynamically stable.

The Kirchhoff model~\cite{kirchhoff} has been extensively used to
model the structure and elasticity of long DNA
chains~\cite{olson,coleman,coleman2,fonseca1,fonseca2,fonseca3,manning},
the tendrils of climbing plants~\cite{alain1,alain2}, slender cables
subject to different stresses~\cite{got}, etc. The Kirchhoff model is
also appropriate for investigating the elastic properties of amorphous
helical nanostructures.

For nanowires with elliptic cross-section, there are two types of
helical structures, {\it normal} and {\it binormal}. In Section II, we
present the Kirchhoff rod model and, for each type of helix, derive
two expressions that can be used to obtain the Young's modulus and the
Poisson's ratio of an amorphous nanospring composite material. One of
expressions relates the Hooke's constant to the Young's modulus, and
to the geometric features of the nanospring. The other expression
relates an applied torque in the direction of the helical axis of the
nanospring, to the Young's modulus and Poisson's ratio of the
material. In Section III, we used the measured Hooke's constant for a
carbon nanocoil, reported in Ref.~\cite{chen}, to test the expression
for the Young's modulus. We discuss the results and analyse the
stability of helical filaments, showing that the presence of a
metallic catalyst has important consequences in the growth process of
nanosprings. In section IV we summarize our results and conclusions.

\section{The elastic model} 

In this section, we derive two expressions that can be used by
experimentalists to obtain the elastic constants of the composite
material of amorphous nanosprings. Both expressions involve two
geometric parameters that define a helical space curve, namely, the
{\it curvature}, $\kappa$, and the {\it torsion}, $\tau$.

We shall first briefly describe a helical curve and its relation to
the curvature and torsion. Then, we present the Kirchhoff rod model,
that is used to derive the above mentioned expressions for studying
the elastic properties of amorphous nanosprings.

\subsection{Helical space curve}

A space curve ${\bf{x}}$ is a {\it helix} if the lines tangent to
${\bf{x}}$ make a constant angle with a fixed direction in space (the
helical axis)~\cite{dirk}.

\begin{figure}[ht] 
  \begin{center}
  \includegraphics[height=60mm,width=30mm,clip]{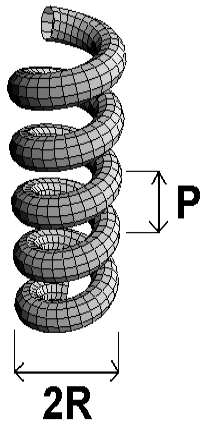}
  \end{center} 
  \caption{A helical rod characterized by a radius $R$ and a loop-to-loop 
  distance $P$.}
\label{fig1}
\end{figure} 

A helical space curve ${\bf x}$ can be expressed in a fixed Cartesian
basis as:
\begin{equation}
\label{helcurve}
{\bf x}=R\cos(\lambda s)\,{\bf e}_x+R\sin(\lambda s)\,{\bf e}_y+
\frac{P}{2\pi}\lambda\, s\, {\bf e}_z \; , 
\end{equation}
where
\begin{equation}
\label{lamb1}
\lambda=\sqrt{\frac{1}{(R^2+\frac{P^2}{4\pi^2})}} \; ,
\end{equation}
$R$ is the radius of the helix, and $P$ is the {\it pitch} of the
helix, i.e, the distance between two adjacent loops. $\{{\bf e}_x,{\bf
e}_y,{\bf e}_z\}$ is a fixed Cartesian basis where ${\bf e}_z$ is
chosen along the direction of the axis of the helix. Fig. \ref{fig1}
shows a helical filament of radius $R$, and pitch $P$.

The radius, $R$, and pitch, $P$, of the helix are related to the
curvature, $\kappa$, and torsion, $\tau$, through:
\begin{equation}
\label{R}
\begin{array}{l}
\kappa=R\lambda^{2} \; , \\
\tau=\frac{P}{2\pi}\lambda^{2} \; .
\end{array}
\end{equation}
In terms of $\kappa$ and $\tau$, the helical curve ${\bf x}$ can be
written as~\cite{alain2}:
\begin{equation}
\label{helcurve2}
{\bf x}=\frac{\kappa}{\lambda^2}\cos(\lambda s)\,{\bf e}_1 +
\frac{\kappa}{\lambda^2}\sin(\lambda s)\,{\bf e}_2 +
\frac{\tau}{\lambda}\,s\,{\bf e}_3 \; ,
\end{equation}
where $\lambda$ can also be written in terms of $\kappa$ and $\tau$
by:
\begin{equation}
\label{lamb2}
\lambda=\sqrt{\kappa^2+\tau^2} \; .
\end{equation}

These equations will be useful to derive the relations for the elastic
constants of the nanospring.

\subsection{The Kirchhoff rod model}

In Kirchhoff's theory, the rod is seen as an assembly of short
segments loaded by contact forces from the adjacent ones. The
classical equations for the conservation laws of linear and angular
momentum are applied to each segment in order to obtain a one
dimensional set of differential equations for the static and dynamics
of the rod in the approximation of large radius of curvature and large
total length of the rod as compared to the radius of the local
cross-section~\cite{dill}. These equations contain the forces and
torques, plus a triad of vectors describing the deformations of the
rod. In this paper, we shall be concerned only with static solutions
and, therefore, only the static Kirchhoff equations will be presented:
\begin{equation}  
\label{Kir1}  
{\bf F}' = 0 \; ,
\end{equation}  
\begin{equation}  
\label{Kir2}  
{\bf M}' + {\bf d}_{3}\times {\bf F} = 0 \; ,
\end{equation}  
where ${\bf F}$ and ${\bf M}$ are the total force and torque across
the cross-sections of the rod, respectively. ${\bf d}_3$ is the vector
tangent to the centerline or the axis of the rod. The prime denotes
the derivative with respect to the arc-length $s$ of the rod. In order
to solve the equations we introduce the constitutive relationship from
linear elasticity theory~\cite{dill} that, for a rod with elliptic
cross-section, is given by~\cite{alain3}:
\begin{equation}
\label{Kir3}
{\bf M}=EI_1(k_1-k^{(0)}_{1}){\bf d}_1+EI_1a(k_2-k^{(0)}_{2}){\bf d}_2
+EI_1b(k_3-k^{(0)}_{3}){\bf d}_3 \; ,
\end{equation}
where $E$ is the Young's modulus, $a\equiv I_2/I_1$, with $I_1$ and
$I_2$ being the principal moments of inertia of the cross section in
the directions of ${\bf d}_1$ and ${\bf d}_2$, respectively, with
$I_1\ge I_2$. Since the bending coefficient of the rod is proportional
to the moment of inertia, the vectors ${\bf d}_1$ and ${\bf d}_2$
represent the directions of greatest and lowest bending stiffness of
the rod, respectively. $a$, $0<a\le1$, measures the bending asymmetry
of the cross section. ${\bf d}_1$ and ${\bf d}_2$ lie in the plane of
the cross section so that $\{{\bf d}_1$,${\bf d}_2$,${\bf d}_3\}$
forms a right handed director basis defined at each point along the
axis of the rod. The constant $b$ is called {\it scaled torsional
stiffness} and for a rod of elliptic cross-section with semiaxes $A$
and $B$ ($A<B$), $a$ and $b$ are given by~\cite{alain3,landau,love}:
\begin{equation}
\label{ab}
a=\frac{A^2}{B^2}\, , \; \; \; b=\frac{1}{1+\sigma}\frac{2a}{1+a} \; .
\end{equation}
where $\sigma$ is the Poisson's ratio of the material.

$k_j$, $j=1,2,3$, are the components of the so-called {\it twist
vector}, ${\bf k}$, which defines the variation of the director basis
$\{{\bf d}_1,{\bf d}_2,{\bf d}_3\}$ with the arc-length $s$ through
the expression:
\begin{equation}
\label{veck} 
{\bf d}'_j={\bf k}\times{\bf d}_j \, , \; j=1,2,3 \, . 
\end{equation}
$k_1$ and $k_2$ are related to the curvature $\kappa$ of the
centerline of rod through $\kappa=\sqrt{k^2_1+k^2_2}$, and $k_3$ is
the twist density of the rod. $k^{(0)}_j$, $j=1,2,3$, defines the
variation of the director basis of the rod in its unstressed
configuration. $k^{(0)}_j$, $j=1,2,3$, represent the {\it intrinsic
curvature} of the rod, which is the tridimensional configuration
displayed by the rod when it is free from stresses.

\begin{figure}[ht] 
  \begin{center}
  \includegraphics[height=60mm,width=30mm,clip]{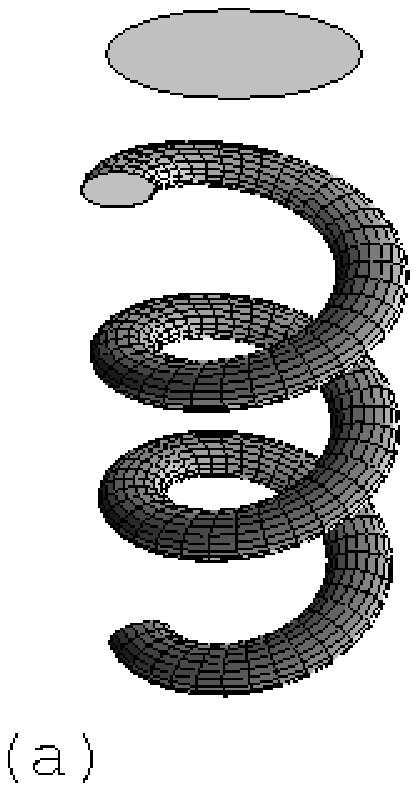}
  \vspace{1cm}
  \includegraphics[height=70mm,width=30mm,clip]{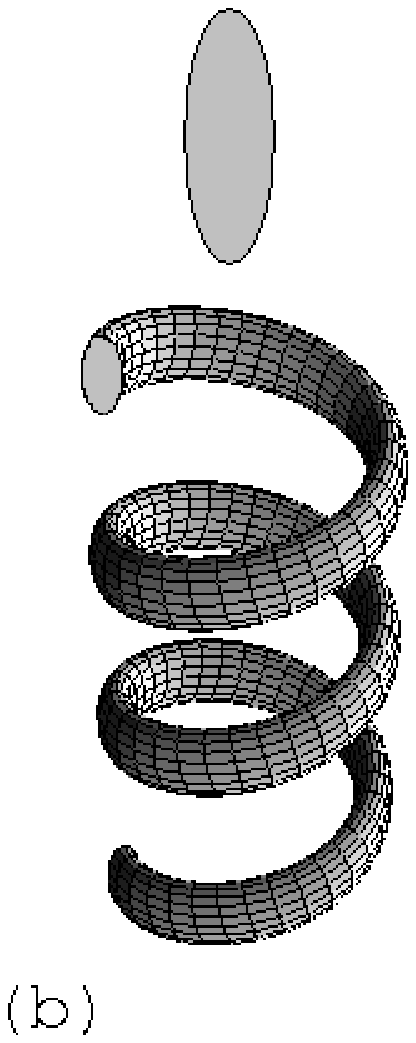}
  \end{center} 
  \caption{{\it Normal} (a) and {\it binormal} (b) helices and the 
  corresponding orientated cross-sections.}
\label{fig2}
\end{figure} 

The Kirchhoff equations (\ref{Kir1}-\ref{Kir3}) admit two types of
helical solutions for rods with elliptic cross section, called {\it
normal} and {\it binormal} helices. A helical solution is represented
by a set of expressions for the twist vector and the force, and the
upper index $n$ ($b$) will be used to denote {\it normal} ({\it
binormal}) helical solutions. Figure \ref{fig2} displays examples of
these types of helix. The {\it normal} helix solution is given by:
\begin{eqnarray}
\label{hnor}
&&{\bf k}^n=\kappa{\bf d}_1 + \tau{\bf d}_3 \; , \label{hn1} \\
&&{\bf F}^n=\gamma_1\tau\,{\bf k}^n \; . \label{hn2}
\end{eqnarray}
The {\it binormal} helix solution is given by:
\begin{eqnarray}
\label{hbi}
&&{\bf k}^b=\kappa{\bf d}_2 + \tau{\bf d}_3 \; , \label{hb1} \\
&&{\bf F}^b=\gamma_2\tau\,{\bf k}^b \; . \label{hb2} 
\end{eqnarray}
where $\kappa$ ({\it curvature}) and $\tau$ ({\it torsion}) are the
geometric parameters of the space curve defined by the helical axis of
the rod, and
\begin{equation}
\label{gammai}
\gamma_i=b(1-\frac{\tau_0}{\tau})-
[1+(a-1)\delta_{i2}](1-\frac{\kappa_0}{\kappa}) \; , \; i=1,2 \; . 
\end{equation}
where $\kappa_0$ and $\tau_0$ are the {\it intrinsic} curvature and
torsion of the helical structure (${\bf k}^{(0)}=\kappa_0{\bf
d}_i+\tau_0{\bf d}_3$), $i=1$ ($i=2$) for a {\it normal} ({\it
binormal}) helix. $\delta {}_{i2}$ is the Kronecker delta.

We can relate the director basis, $\{{\bf d}_1,{\bf d}_2,{\bf d}_3\}$,
to the fixed Cartesian basis, $\{{\bf e}_x,{\bf e}_y,{\bf e}_z\}$,
integrating the eq. (\ref{veck}) using the eqs. (\ref{hn1}) and
(\ref{hb1}) for {\it normal} and {\it binormal} helices,
respectively. We obtain the following relations:
\begin{eqnarray}
\label{dn}
&&{\bf d}^{n}_1=\frac{1}{\lambda}(\tau\sin(\lambda s){\bf e}_x -
\tau\cos(\lambda s){\bf e}_y + \kappa\,{\bf e}_z) \; , \label{dn1} \\
&&{\bf d}^{n}_2=\cos(\lambda s){\bf e}_x +
\sin(\lambda s){\bf e}_y\; , \label{dn2} \\
&&{\bf d}^{n}_3=\frac{1}{\lambda}(-\kappa\sin(\lambda s){\bf e}_x +
\kappa\cos(\lambda s){\bf e}_y + \tau\,{\bf e}_z) \; , \label{dn3} 
\end{eqnarray}
for the {\it normal} helix, and
\begin{eqnarray}
\label{db}
&&{\bf d}^{b}_1=\sin(\lambda s){\bf e}_x +
\cos(\lambda s){\bf e}_y  \; , \label{db1} \\
&&{\bf d}^{b}_2=\frac{1}{\lambda}(\tau\cos(\lambda s){\bf e}_x -
\tau\sin(\lambda s){\bf e}_y + \kappa\,{\bf e}_z) \; , \label{db2} \\
&&{\bf d}^{b}_3=\frac{1}{\lambda}(-\kappa\cos(\lambda s){\bf e}_x +
\kappa\sin(\lambda s){\bf e}_y + \tau\,{\bf e}_z) \; , \label{db3} 
\end{eqnarray}
for the {\it binormal} helix. Since ${\bf d}_3={\bf x}'$ we can
integrate eqs. (\ref{dn3}) and (\ref{db3}) to obtain an expression for
${\bf x}$ similar to the eq. (\ref{helcurve2}), except for a
difference in phase.

From eqs. (\ref{hn2}) and (\ref{hb2}) we can obtain the total tension
force $T$ along the direction of the axis of the helix (here defined
as ${\bf e}_z$):
\begin{equation}
\label{tension}
T={\bf F}.{\bf e}_z=\gamma_i\tau{\bf k}.{\bf e}_z \; , \; i=1,2 \; ,
\end{equation}
where $\gamma_i$ is given by eq. (\ref{gammai}) and $i=1$ ($i=2$) is
related to the vectors ${\bf F}$ and ${\bf k}$ for the {\it normal}
({\it binormal}) helix given by eqs. (\ref{hnor})
(eqs. (\ref{hbi})). Since the twist vector ${\bf k}$ is written in the
director basis through eq. (\ref{hn1}) (eq.  (\ref{hb1})), we can use
eqs. (\ref{dn}) (eqs. (\ref{db})) to obtain $T$:
\begin{equation}
\label{T}
T=\frac{EI_1}{L^2}\gamma_i\,L^2\tau\,\lambda=EI_1\,\gamma_i\,\tau\,\lambda 
\; , \; i=1,2 \; ,
\end{equation}
where the terms $EI_1/L^2$ and $L^2$ appeared in accord to the
following conversion to unscaled variables:
\begin{equation}
\label{convert}
T\rightarrow\frac{L^2}{EI_1}T \; \; , \; \; 
\kappa\rightarrow L\kappa \; \; , \; \; \tau\rightarrow L\tau
\; \; \mbox{and} \; \;  \lambda\rightarrow L\lambda \; ,
\end{equation}
with $L$ being a length scale of the rod that, in our case, is chosen
to be the arc-length of one loop of the unstressed helical rod:
$L=2\pi/\lambda_0$ with $\lambda_0=\sqrt{\kappa^2_0+\tau^2_0}$. $E$ is
the Young's modulus of the composite material of the rod, and $I_1$ is
the moment of inertia along ${\bf d}_1$ (it is the direction of the
greatest bending stiffness). For an elliptic cross section of semiaxes
$A$ and $B$ ($A<B$), $I_1$ is given by:
\begin{equation}
\label{I1}
I_1=\frac{B^3A\pi}{4} \; . 
\end{equation}
%

\subsection{Expressions for the elastic constants of the nanospring}

McMillen and Goriely~\cite{alain2} used the Kirchhoff model to obtain
an expression for the Hooke's constant $h$ of a helix, with intrinsic
curvature ${\bf k}^{(0)}=\kappa_0{\bf d}_1+\tau_0{\bf d}_2$, in terms
of the properties of the rod's material, the Young's modulus, and the
moment of inertia of the circular cross-section.

Here, we follow the same steps to derive an expression for the Hooke's
constant of a helical filament with elliptic cross section. We also
derive an expression relating a torque applied along the direction of
the helical axis, to the Poisson's ratio of the helical filament.

Consider a helical rod in its unstressed state represented by ${\bf
k}^{(0)}=\kappa_0{\bf d}_i+\tau_0{\bf d}_3$, $i=1$ ($i=2$) for {\it
normal} ({\it binormal}) intrinsically helical configuration. The ends
of the rod are held fixed so that they do not rotate as the applied
tension at the ends is changed. Since the ends do not rotate, it
imposes a constraint in the total twist $T_W$ of the rod:
\begin{equation}
\label{tw}
T_W\equiv\int k_3\,ds=\int \tau_0\,ds=\int \tau\,ds \; . 
\end{equation}
The consequence of this constraint is that the torsion $\tau$ remains
constant ($\tau=\tau_0$) for the tensioned helical rod.

Consider two material points, $Q_1$ and $Q_2$, in the unstressed
configuration, that are located at arc-length positions $s_1=0$ and
$s_2=2\pi/\lambda_0$, respectively. $2\pi/\lambda_0$ is exatcly the
arc-length of one loop of the unstressed helical configuration. When a
tension is applied to the rod, the distance between those points $Q_1$
and $Q_2$, along the helix axis, is given by
$d=z(2\pi/\lambda_0)-z(0)$.  Using eq. (\ref{helcurve2}), $d$ can be
written as
\begin{equation}
\label{distance}
d=\frac{2\pi\tau}{\lambda \lambda_0} \; ,
\end{equation}
where $\lambda$ and $\tau$ are related to the deformed (stretched or
compressed) helical structure. From (\ref{distance}), we obtain the
expression for $\tau$:
\begin{equation}
\label{tau3}
\tau=\tilde{d}\lambda \; , 
\end{equation} 
where  $\tilde{d}$ is the dimensionless distance
\begin{equation}
\label{dtil}
\tilde{d}\equiv\frac{d}{L}=d\frac{\lambda_0}{2\pi} \; . 
\end{equation}
The eq. (\ref{gammai}) can be simplified for this case where
$\tau=\tau_0$:
\begin{equation}
\label{gammaiS}
\gamma_i=[1+(a-1)\delta_{i2}]\left(\frac{\kappa_0}{\kappa}-1\right) \; , \; 
i=1,2 \; .
\end{equation}
Substituting eq. (\ref{gammaiS}) in eq. ({\ref{T}) we obtain the
following expression for the tension $T$:
\begin{equation}
\label{T2}
T=EI_1[1+(a-1)\delta_{i2}]\left(\frac{\kappa_0}{\kappa}-1\right)\tau 
\lambda \; , \; i=1,2 \; . 
\end{equation}
From the relation (\ref{lamb2}), using eq.~(\ref{tau3}), we obtain
\begin{equation}
\label{kappa2}
\kappa=\sqrt{\lambda^2-\tau^2}=\lambda\sqrt{1-\tilde{d}^2} \; .
\end{equation}
Substituting  eq.~(\ref{kappa2}) in eq.~(\ref{T2}), and using 
eq.~(\ref{tau3}), we obtain an expression for $T$ as function of
$\tilde{d}$:
\begin{equation}
\label{T3}
T=EI_1[1+(a-1)\delta_{i2}]\left(\frac{\kappa_0\tau}{\sqrt{1-\tilde{d}^2}}-
\frac{\tau^2}{\tilde{d}}\right)
\; , \; i=1,2 \; .
\end{equation}
If $\tilde{d}_0$ is the distance between the points $Q_1$ and $Q_2$ in
the unstressed configuration of the rod, then
$\tilde{d}=\tilde{d}_0=\tau_0/\lambda_0$ when no tension is applied to
the rod. Eq (\ref{T3}) shows that the tension $T$ varies from $0$ to
$\infty$ if $\tilde{d}$ varies from $\tilde{d}_0$ to $1$. In order to
find the Hooke's constant, $h$, of the tensioned helical rod, we
consider a small variation of $\tilde{d}$ defined by the small
displacement $\hat{d}$, i.e., we let $\tilde{d}=\tilde{d}_0+\hat{d}$,
$\hat{d}\ll1$. Then we look for an expression for the Hooke's constant
such that:
\begin{equation}
\label{Th}
T=h\hat{d}+O(\hat{d}^2) \; . 
\end{equation}
Expandind eq. (\ref{T3}) for a small variation of $\tilde{d}$ around
$\tilde{d}_0$, we obtain
\begin{equation}
\label{T4}
T=EI_1[1+(a-1)\delta_{i2}]\lambda^2_0
\left(1+\frac{\tau^2_0}{\kappa_0^2}\right)\,\hat{d} + O(\hat{d}^2)
\; , \; i=1,2 \; .
\end{equation}
For a spring with $N$ coils, its displacement due to the action of a
given tension $T$ is obtained replacing $\{\hat{d}\}$ by
$\left\{\frac{\hat{d}}{NL}\right\}$ in the above equation where, now,
$\hat{d}$ is written in unscaled units. It gives the following final
expression for the Hooke's constant, $h$:
\begin{equation}
\label{hooke}
h=\frac{EI_1}{2\pi N}[1+(a-1)\delta_{i2}]\lambda^3_0
\left(1+\frac{\tau^2_0}{\kappa_0^2}\right) \; , \; i=1,2 \; ,
\end{equation}
where $i=1$ ($i=2$) refers to the {\it normal} ({\it binormal}) helix
solution.  By making $a=1$ we recover our previous
result~\cite{douglas} for the Hooke's constant of a nanospring made of
a nanowire with circular cross section.

Eq.~(\ref{hooke}) relates the Hooke's constant, $h$, to the Young's
modulus, $E$, of the material composite of the nanospring, to the
moment of inertia, $I_1$, and to the geometric parameters $\kappa_0$,
$\tau_0$ and $\lambda_0=\sqrt{\kappa^2_0+\tau^2_0}$. The
moment of inertia, $I_1$, and the parameter $a$ can be obtained by
measuring the semiaxes $A$ and $B$ of the elliptic cross section of
the nanowire and using Eqs.~(\ref{I1}) and (\ref{ab}),
respectively. The geometric parameters can be obtained by measuring
the radius, $R$, and the pitch, $P$, of the nanospring and using the
eqs.~(\ref{lamb1}) and (\ref{R}) that relates $R$ and $P$ to
$\lambda$, $\kappa$ and $\tau$. To obtain the values of the geometric
paramaters of the unstressed configuration, we must measure $R_0$ and
$P_0$ of the unstressed helical configuration, and then use the
eqs.~(\ref{lamb1}) and (\ref{R}) to obtain $\lambda_0$, $\kappa_0$ and
$\tau_0$. Therefore, by measuring the Hooke's constant of the
nanospring we can use eq.~(\ref{hooke}) to obtain the Young's modulus
of the composite material of the nanospring. Figure~\ref{fig3} shows a
scheme that can be used for measuring the Hooke's constant of the
nanospring.

\begin{figure}[ht] 
  \begin{center}
  \includegraphics[height=60mm,width=120mm,clip]{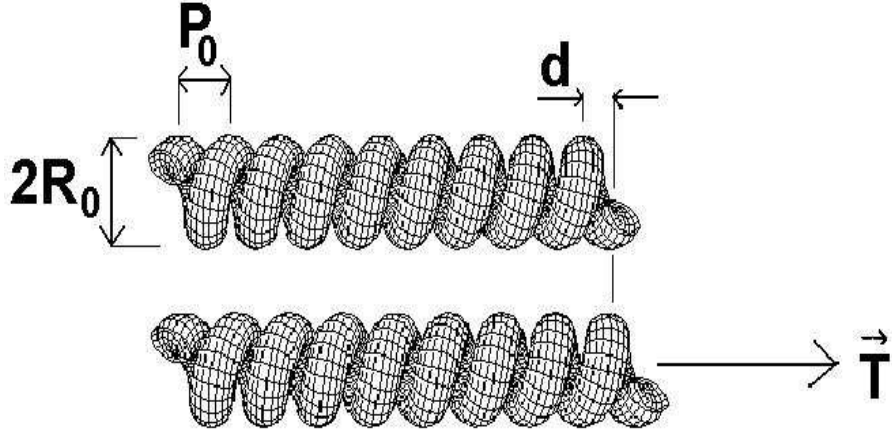}
  \end{center} \caption{Outline of the experiment to measure the
  Hooke's constant, $h$. $R_0$ and $P_0$ are the radius and the pitch of the 
  helix in the unstressed configuration. The helix shown corresponds 
  to a nanospring of radius $R_0=51$nm and $P_0=85$nm.}
\label{fig3}
\end{figure} 

We, now, derive an expression to obtain the Poisson's ratio $\sigma$
of the nanospring. We depart from the eq.~(\ref{Kir3}) for the total
torque across each cross section, and then use eq.~(\ref{hn1})
(eq.~(\ref{hb1})) for the components of the twist vector ${\bf k}$ for
a {\it normal} ({\it binormal}) helix. Basically, we want to obtain
the axial component of the torque $M_Z\equiv{\bf M}.{\bf e}_z$. Using
eqs.~(\ref{dn}) and (\ref{db}) we obtain the following expression for
$M_Z$:
\begin{equation}
\label{MZ}
M_Z=[1+(a-1)\delta_{i2}](\kappa-\kappa_0)\frac{\kappa}{\lambda}+
b(\tau-\tau_0)\frac{\tau}{\lambda} \; , \; i=1,2 \; ,
\end{equation}
where $i=1$ ($i=2$) corresponds to a {\it normal} ({\it binormal})
helical nanospring.

Eq.~(\ref{MZ}) can be used to obtain, experimentally, the Poisson's
ratio $\sigma$ of the nanospring. Figure~\ref{fig4} shows a scheme to
obtain the Poisson's ratio. Measuring the radius, $R_1$, and the
pitch, $P_1$, of the stressed helix, $\kappa$ and $\tau$ can be
obtained from eqs.~(\ref{lamb1}) and (\ref{R}). By measuring the
applied torque in the direction of the axis of the helix, $M_Z$, we
can use the Eq. (\ref{MZ}) to obtain the parameter $b$. The values of
$a$, $\kappa_0$, $\tau_0$, $I_1$ and the nanospring Young's modulus
are needed to obtain $b$. The Poisson's ratio, $\sigma$, can,
therefore, be obtained using the Eq. (\ref{ab}). In this experimental
scheme, both ends of the nanospring must be held fixed in order to
avoid relaxation.

\begin{figure}[ht] 
  \begin{center}
  \includegraphics[height=70mm,width=130mm,clip]{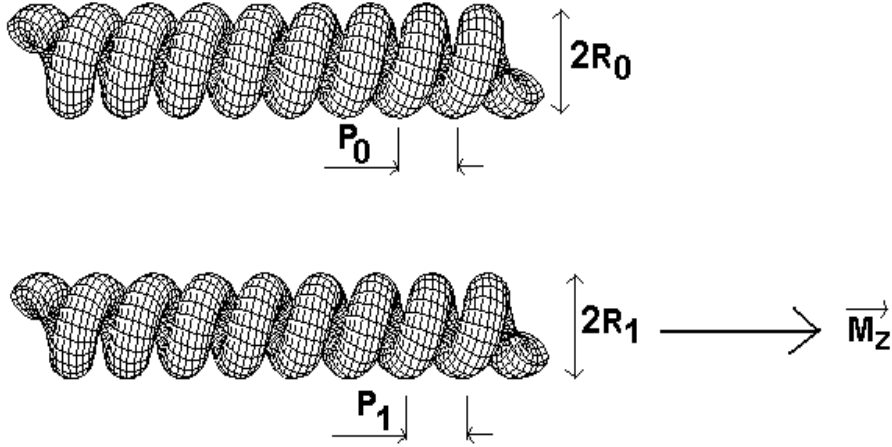}
  \end{center} \caption{Outline of the experiment to measure the
  Poisson's ratio $\sigma$. The extremities of the nanospring must be
  held fixed. See the text for the details.}
\label{fig4}
\end{figure} 


\section{Results and discussions}

In this section, we test and discuss the consequences of our model
(Eqs. (\ref{hooke}) and (\ref{MZ})), and show the importance of the
presence of the catalytic particle in the growth process of
nanosprings.

\subsection{Test and discussion of the model}

Equation (\ref{hooke}) can be tested using the parameters for a unit
nanocoil ($N=1$) considered by Chen {\it et al}~\cite{chen}: the
diameter of the nanowire $D=120$nm, the radius of the helix $R=420$nm
and the pitch of the helix $P=2000$nm. The material has a Poisson's
ratio $\sigma=0.27$ and a shear modulus $\mu=2.5$GPa. The amorphous
carbon nanocoil used by Chen {\it et al} is, in fact, a double coil
formed by two wires tightly joined~\cite{chen,huang}, as indicated by
scanning electron microscope (SEM) and transmission electron
microscope (TEM) images depicted in Figure 7 of Reference
\cite{chen}. To test our model, we shall approximate the carbon 
nanocoil by a homogeneous rod with circular cross section of diameter
$D$.

Assuming that the rod has circular cross section, eq.~(\ref{ab}) gives
$a=1$ and $b=(1+\sigma)^{-1}=2\mu/E$. Using the values given above for
$\mu$ and $\sigma$, we obtain that the nanocoil material has the
Young's modulus $E_{bulk}=6.35$GPa.

Now, using the equations (\ref{lamb1}) and (\ref{R}) we obtain
$\kappa_0$ and $\tau_0$ of the carbon nanocoil from the values of $R$
and $P$ given above. Then, using the measured Hooke's constant of the
nanocoil, $h=0.12$N/m~\cite{chen}, the Eq.~(\ref{hooke}), with $a=1$
(circular cross section), and $N=1$ (unit nanocoil), gives
$E_{nanocoil}\simeq6.88$GPa, which is slightly larger than
$E_{bulk}=6.35$GPa. 

This difference could not be only due to the nanocoil having
non-circular cross section. We should point out that experimentalists
have observed that the measured Young's modulus and Poisson's ratio of
nanostructures differ from those of the bulk material. For instance,
Salvadori {\it et al} found that the Young's modulus of gold thin film
is about $12\%$ smaller than that of the bulk
material~\cite{salva,salva2}. Using alternating electrical fields,
Dikin {\it et al}~\cite{dikin} excited bending vibrations in SiO$_2$
nanowires to measure their Young's modulus, obtaining values smaller
than that of the bulk SiO$_2$ material. In another example, Cuenot
{\it et al}~\cite{cuenot} used the effects of surface tension in bent
structures to explain the larger values of the Young's modulus
observed in nanowires with smaller diameters. Our model allows to
obtain the Young's modulus and the Poisson's ratio of a helical
nanostructure directly, it does not use the bulk material values.

Eq. (\ref{hooke}) shows that, for a nanowire where $a\neq1$ (non
symmetric cross section, Eqs. (\ref{Kir3}) and (\ref{ab})), the
Hooke's constant of the {\it normal} helical configuration is always
larger than that of the {\it binormal} one. It indicates that it is
possible to produce helical nanowires whose axes have the same
curvature and torsion but different elastic properties, what is useful
in nanoengineering.

Eq~(\ref{MZ}) cannot be tested due to lack of experimental
measurements of the axial torque of a nanospring. However, the
recently developed method of measuring axial torque applied to
polymers~\cite{oro} could help the development of experiments to apply
axial torques to nanosprings and nanowires. Eq. (\ref{MZ}) shows that,
for a nanowire where $a\neq1$, the torque required to twist a
nanospring is different for {\it normal} and {\it binormal}
configurations.

\subsection{Mechanical function of the catalyst} 

McIlroy {\it et al}~\cite{mc1,mc2} have proposed a model of nanospring
formation based upon the vapour-liquid-solid (VLS) growth mechanism by
Wagner and Ellis~\cite{ellis}. The key feature of the VLS model is the
presence of a liquid catalyst that absorbs the material from
surrouding vapour and, after becoming supersaturated, the material is
deposited beneath the catalyst-substrate interface, thereby forming
the nanowire~\cite{mc1,mc2}.

The geometry of the catalyst has a strong influence on the geometry of
the growing nanowire and this feature was explored by McIlroy {\it et
al} in their modified VLS growth model for nanosprings~\cite{mc1,mc2}.
Since the growth is driven by the interaction of the surface tensions
between the liquid-vapour ($\gamma_{LV}$), solid-vapour
($\gamma_{SV}$) and solid-liquid ($\gamma_{SL}$) interfaces, they
proposed that the asymmetric growth of the nanowire, which leads to
the helical shape of the nanostructure, occurs due to the contact
angle anysotropy (CAA) at the catalyst-nanowire interface. The work
required to shear the catalyst from the nanowire is called the
thermodynamic work of adhesion $W_A$ and can be computed in terms of
the surface tensions by~\cite{mc1}:
\begin{equation}
\label{wa}
\begin{array}{ccc}
W_A&=&\gamma_{SV}+\gamma_{SL}-\gamma_{LV} \; \\
&=&\gamma_{SV}(1-\cos\theta) 
\end{array}
\end{equation}
where $\theta$ is the angle between the surface tensions $\gamma_{SL}$
and $\gamma_{SV}$. 

We shall show that, from the mechanical point of view, the catalytic
particle has an important role in the stability of the grown
nanosprings through the VLS mechanism. We also show that any other
mechanism of growth that produces {\it differential growth} can
produce stable helical nanostructures.

Our analysis depart from the mechanical conditions for the stability
of helical structures. 

Goriely and Tabor developed a dynamical method to test the stability
of equilibrium solutions of the Kirchhoff model~\cite{alain4,alain42},
and showed~\cite{alain5} that helices with intrinsic curvature
($\kappa_0\neq0$ and $\tau_0\neq0$) do not admit unstable modes being,
therefore, dynamically stable. Therefore, {\it intrinsic curvature} is
the key feature for stability of helices and it comes just from the
{\it differential growth} of a one-dimensional structure.

McMillen and Goriely~\cite{alain2} have pointed out that the {\it
differential growth} in tendrils of climbing plants produces {\it
intrinsic curvature}. According to the McIlroy's modified VLS model,
the CAA at the catalyst-nanowire interface is responsible for the
asymmetric growth of the forming helical nanowire. Therefore, the CAA
{\it induces a helical differential growth}, thus producing the {\it
helical intrinsic curvature} required for dynamical stability of the
forming nanospring. 

The importance of the catalyst in the {\it differential growth} of the
helical nanowire is that the CAA depends on the presence of the
catalyst. Therefore, we can conclude that the catalytic particle is
responsible for conferring the mechanical stability on the nanosprings
grown by the VLS growth model. It shows that the McIlroy {\it et al}
VLS growth model is consistent with the mechanical stability of the
grown helical nanostructures.

This importance is corroborated by the analysis of a recent report on
spontaneous polarization-induced growing of nanosprings and nanorings
of piezoeletric zinc oxide (ZnO) by Kong and Wang~\cite{wang}. Their
crystalline nanosprings were produced without the presence of
catalytic particles. They found that the mechanism for the formation
of the helical structure is the consequence of the interplay between
the electrostatic forces between the polarized material and the
substract, and the elastic forces that holds the nanostructure in the
helical conformation. If these helical nanostructures were moved away
from the substract, the electrostatic forces would cease and the
nanostructure would release its elastic energy becoming straight,
implying that these nanostructures do not have intrinsic helical
curvature. The growth process of ZnO nanohelices does not produce {\it
intrinsic curvature}. 

Some nanowires grown by Chemical Vapour Deposition (CVD) are not
assisted by a {\it liquid} catalytic particle, indicating that the
growth of such nanowires occurs by a Vapour-Solid-Solid (VSS)
mechanism~\cite{lars2,lars3}. We have given emphasis to the VLS growth
mechanism because there are several examples of nanospring growth
explained by the VLS mechanism~\cite{mc1,mc2,zhang,mc3}. It should be
stressed that our analysis can be extended to any type of growth
mechanism (including VSS) provided that the growth process presents
the geometric features necessary to create the {\it differential
growth}.

\section{Conclusions}

The Kirchhoff rof model has been used to obtain a simple model to
investigate the elastic properties of amorphous nanosprings. We
derived the expressions (\ref{hooke}) and (\ref{MZ}) that can be used
to obtain the Young's modulus and the Poisson's ratio of the composite
material of a nanospring. These expressions relate the material
elastic constants to the geometric features of the nanowire helical
structure, and cross section. 

We have considered the case of nanowire with elliptic cross section
that leads to two types of helical solutions for the Kirchhoff
equations: {\it normal} and {\it binormal} helices
(Fig. \ref{fig2}). In the particular case of nanowire with circular
cross section, we recover the expressions derived in
Ref.~\cite{douglas}.

We also proposed the schemes (figures \ref{fig3} and \ref{fig4}) to be
used for measuring the Hooke's constant, and the applied torque along
the direction of the axis of the helix, together with the radius, $R$,
and the pitch, $P$, of the helical structure.  Eqs. (\ref{lamb1}) and
(\ref{R}) can be used to obtain the curvature, $\kappa$, and torsion,
$\tau$, of the nanospring, necessary to use Eqs. (\ref{hooke}) and
(\ref{MZ}) to obtain the Young's modulus and the Poisson's ratio of
the nanospring.

We showed that the presence of the catalyst is important for
conferring the dynamical stability in the nanosprings grown by VLS
mechanism. We showed that the {\it differential growth} produced by
the asymmetries in the growth process of a helical nanowire is
responsible for the production of its {\it intrinsic curvature}. We
hope our analysis estimulate experimentalists to investigate more
features of the growth processes of helical nanostructures.

\ack This work was partially supported by the FAPESP, CNPq, IMMP, IN,
SAMNBAS and FINEP.

\end{document}